# Universal point spread function engineering for 3D optical information processing

Md Sadman Sakib Rahman[1,2,3]    mssr@ucla.edu
Aydogan Ozcan[1,2,3,*]            ozcan@ucla.edu

[1]Electrical and Computer Engineering Department, University of California, Los Angeles, CA, 90095, USA
[2]Bioengineering Department, University of California, Los Angeles, CA, 90095, USA
[3]California NanoSystems Institute (CNSI), University of California, Los Angeles, CA, 90095, USA
[*]Corresponding author: ozcan@ucla.edu

## Abstract

Point spread function (PSF) engineering has been pivotal in the remarkable progress made in high-resolution imaging in the last decades. However, the diversity in PSF structures attainable through existing engineering methods is limited. Here, we report universal PSF engineering, demonstrating a method to synthesize an arbitrary set of spatially varying 3D PSFs between the input and output volumes of a spatially incoherent diffractive processor composed of cascaded transmissive surfaces. We rigorously analyze the PSF engineering capabilities of such diffractive processors within the diffraction limit of light and provide numerical demonstrations of unique imaging capabilities, such as snapshot 3D multispectral imaging without involving any spectral filters, axial scanning or digital reconstruction steps, which is enabled by the spatial and spectral engineering of 3D PSFs. Our framework and analysis would be important for future advancements in computational imaging, sensing and diffractive processing of 3D optical information.

## Introduction

The spreading or blurring of a point source of light by an optical system is described by its point spread function (PSF)[1]. PSF engineering[2], which involves the purposeful design of the PSF of an optical system, offers a powerful tool for optical imaging and microscopy[3–6]. For example, 3D localization microscopy has greatly benefitted from PSF engineering, enabling remarkable precision in emitter localization[7–10]. PSF engineering is also important for optical data storage and the design of telescopes[11,12], among many other applications. Therefore, the ability to intricately manipulate or optimize PSFs holds great promise for improvements in the design of optical systems. PSF engineering is usually implemented by placing an appropriately designed phase mask at the pupil (Fourier) plane, which results in a laterally invariant PSF whose functional form remains the same (ideally) as the emitting point source moves laterally over the object plane. However, the spectral and axial variations of such PSFs have been utilized for improved performance in 3D localization microscopy[10,13,14]. An optical system that can also control and engineer laterally varying PSFs, matching a desired set of functions, could provide new degrees of freedom for better imaging performance, especially for task-specific[15–18] imaging and sensing systems.

Diffractive optical processors, comprising successive optimized diffractive surfaces that modulate the amplitude and/or the phase of the incident light waves[19], have emerged as a powerful tool for passive manipulation of light[20]. In this framework, the task-specific optimization of the spatial distribution of diffractive features over the constituent surfaces is performed using deep learning tools on a digital forward model. Following this optimization, the resulting surfaces are fabricated and assembled to form the physical diffractive device, which performs its intended task all-optically through passive light-matter interactions as the input light propagates through a thin volume, typically spanning only a few hundred wavelengths. Such diffractive processors, also known as diffractive optical networks or diffractive networks, have been used for diverse applications ranging from all-optical classification to phase-imaging, optical encryption/decryption, display and 3D imaging, among others[19,21–26]. Capable of performing universal linear transformations, such diffractive processors have also been shown to synthesize arbitrarily chosen spatially and spectrally varying 2D PSFs between the input and output planes[27–30]; however, there have been no studies on 3D PSF engineering using diffractive processors, which is highly important in the context of 3D optical information processing.

Here, we report universal 3D PSF engineering with spatially incoherent diffractive processors, showing that such optical processors can synthesize an arbitrarily defined (desired) set of 3D diffraction-limited PSFs between the voxels of an input volume and an output volume, given that sufficient design degrees of freedom (i.e., diffractive features) are available for optimization. We analyze the effect of diffraction limit on the 3D information processing capacity of diffractive networks and numerically demonstrate a novel application of 3D PSF engineering, namely, snapshot 3D imaging without any digital postprocessing. We also demonstrate *spectrally and spatially* varying 3D PSF engineering for snapshot multispectral 3D imaging – also without the need for any digital postprocessing.

The unique contributions of this work include: (1) the demonstration of universal linear processing of 3D optical information using spatially incoherent monochrome diffractive optical networks, accurately performing any arbitrary (desired) set of spatially varying 3D PSFs; (2) the demonstration of spectrally multiplexed 3D optical information processing with spatially and spectrally programmed 3D incoherent PSFs; and (3) the application of such spatially and spectrally varying 3D incoherent PSFs for snapshot 3D multispectral imaging of a volume of emitters using a single output detector array, i.e., without any spectral filters, axial scanning or digital reconstruction steps needed. Our analyses and results are significant for future developments in computational imaging, sensing and diffractive processing of 3D optical information, such as optical data storage and 3D microscopy.

## Results

In this Article, we use the terms 'diffractive processor' and 'diffractive network' interchangeably. Figure 1 presents the schematic of a spatially incoherent diffractive network that synthesizes an arbitrary set of spatially varying 3D PSFs, defined between the input and output volumes. The input volume is assumed to comprise $C_i$ discrete planes, each of which is discretized into $H_i \times W_i$ diffraction-limited ($\sim \lambda/2$) pixels. The vector $\boldsymbol{i} \in \mathbb{R}_{\geq 0}^{N_i}$ represents the distribution of emission intensities over the $N_i = C_i \times H_i \times W_i$ input voxels. The output volume is similarly discretized into $N_o$ voxels, the intensities of which are represented by the vector $\hat{\boldsymbol{o}}$. The diffractive processor is optimized to create an arbitrary set of $N_i$ 3D PSFs, represented by the $N_i$ columns of the matrix $\boldsymbol{A}$ (see the bottom-right panel of Fig. 1) so that for a given input $\boldsymbol{i}$ the output intensity distribution $\hat{\boldsymbol{o}} = \widehat{\boldsymbol{A}}\boldsymbol{i} \approx \boldsymbol{A}\boldsymbol{i}$. Here, $\widehat{\boldsymbol{A}}$ denotes the all-optical intensity transformation (within a scalar factor) performed by the spatially incoherent diffractive processor. In other words, the diffractive network $\widehat{A}$ is optimized to all-optically approximate an arbitrarily defined linear transformation $A$ between the intensity distributions over the input and output volumes, i.e., $\widehat{A} \approx A$ is approximated for a given set of desired spatially varying incoherent PSFs specified by $A$. In our analysis, we assume spatially incoherent input light, emitted by independent light emitters (for example, fluorescent molecules) distributed within the input volume; we further assume that different point emitters at the input volume do not interact with each other and do not cause shadowing, blocking or secondary excitation of each other. Stated differently, emitter-to-emitter or object-to-object interactions within the input volume are ignored, making the spatially incoherent 3D system linear in intensity (see the 'Discussion' and 'Methods' sections for further details).

First, we examine the all-optical linear transformation errors, i.e., 3D PSF-approximation errors, of our diffractive processors as a function of the number ($N$) of optimizable phase-only diffractive features available within the optical processor. Figure 2a depicts the target (desired) linear transformation $A$, together with an example pair of input ($i$) and target ($o$) intensity distributions satisfying $o = Ai$. The columns of $A$ represent the target 3D PSFs (vectorized). The elements of $A$ are randomly and independently sampled from uniform distributions between 0 and 1 – mimicking any arbitrary set of $N_i$ spatially varying 3D PSFs, where each function is real-valued and non-negative, representing a unique/desired PSF connecting an input voxel to the output voxels. In Fig. 2b, we plot the *transformation errors*, i.e., the errors between the target transformation $A$ and the all-optical transformations $\widehat{A}$ performed by the optimized diffractive processors with $N$ diffractive features distributed over $K$ successive surfaces. To clarify, each point on the interpolated curves of Fig. 2b represents a separately trained diffractive processor with the associated ($K, N$) design value. From the $K = 4$ and $K = 8$ curves, we can see that the transformation error between $A$ and $\widehat{A}$ decreases rapidly to a negligible value as $N$ approaches $2N_iN_o$; beyond $N = 2N_iN_o$ the error does not decrease further. We also note that shallower diffractive networks with, e.g., $K = 2$ successive surfaces have larger errors in approximating the target linear transformation, even when $N$ exceeds $2N_iN_o$, showing the importance of *depth* in a diffractive network architecture[27–30]. These analyses indicate that, given sufficient degrees of freedom $N$ distributed over a deeper architecture, diffractive networks can approximate arbitrary linear transformations between the input and output volumes with negligible error. Figures 2c and 2d further outline the performance of two $K = 4$ diffractive processors with $N \approx 2N_iN_o$ and $N = 4N_iN_o$, labeled as $D_1$ and $D_2$ in Fig. 2b, respectively. For both of these designs, the all-optical linear transformation $\widehat{A}$ representing the 3D PSFs are shown, together with the elementwise absolute differences with respect to the desired transformation, i.e., $|A - \widehat{A}|$, which indicate that the errors in the approximate 3D PSFs are negligible, as desired. These figures also show the respective output 3D intensity distributions ($\widehat{o}$) for the input intensity distribution $i$ depicted in Fig. 2a. The corresponding absolute differences (elementwise) with respect to the target $o$ (shown in Fig. 2a), i.e., $|o - \widehat{o}|$ also reveal negligible errors in the output 3D intensities, further supporting our conclusions.

For a deeper diffractive processor architecture to achieve an accurate all-optical transformation, i.e., $\widehat{A} \approx A$, the empirical convergence threshold required for $N$ is dictated by the number of diffraction-limited voxels at the input ($N_i$) and the output ($N_o$) volumes, and the factor of 2 in this threshold, $2N_iN_o$, is due to the fact that phase-only diffractive features are used as part of the diffractive processor design. While complex-valued optimizable diffractive features would be ideal from the perspective of a 2-fold increase in the independent degrees of freedom available for a given design, phase-only feature-based processors are easier to fabricate and would present lower losses.

Next, we explore the effect of the diffraction limit of light on the approximation of arbitrary 3D PSFs by diffractive networks. We consider two parameters related to the diffraction limit: (1) plane-to-plane distance ($d_{pp}$) within the input and output volumes, and (2) the distances of the farthest planes within the input and output volumes from the first and the last diffractive surfaces ($d_i$ and $d_o$), which dictate the input and output numerical apertures (NA) of the diffractive processor, respectively. Without loss of generality, we assume a symmetric architecture where $d_i = d_o$ and the interplane distances $d_{pp}$ within the input and output volumes are the same; see the top panel of Fig. 3. As for the diffractive processor architecture, we assume $N = 4N_iN_o$ optimizable phase-only diffractive features distributed over $K = 4$ surfaces. In Fig. 3a, we show the trends of the transformation error as a function of $d_{pp}$, parameterized by the corresponding values of $d_i = d_o$. For a given $d_i = d_o$ value, e.g., $10.5\lambda$, the transformation error increases monotonically as $d_{pp}$ decreases beyond the axial resolution set by the input/output NA of the processor. For a larger $d_i = d_o = 21.0\lambda$ (i.e., a smaller NA), the axial resolution of the diffractive processor is even more limited, and the onset of the monotonic error increase occurs at a larger $d_{pp}$ value compared to the $d_i = d_o = 10.5\lambda$ case – as expected. This dependence on NA can be further explained by plotting the transformation error as a function of $d_i = d_o$ for a given $d_{pp}$, as shown in Fig. 3b. For $d_{pp} = 3.0\lambda$, the transformation error increases rapidly as $d_i = d_o$ increases beyond $21.0\lambda$, which results in an axial resolution limit larger than $3\lambda$ (i.e., $\frac{2\lambda}{\text{NA}^2}$=~$3.05\lambda$ where NA $\approx 0.81$ for $d_i = d_o = 21\lambda$ and a diffractive layer width of $W \approx 57\lambda$; see the Methods section).

The ability of diffractive processors to arbitrarily process the 3D intensity information of an input volume through spatially varying 3D PSFs can give rise to interesting applications in computational imaging. Next, we focus on the numerical demonstration of one such application, i.e., the snapshot 3D imaging of independent emitters distributed over a volume. This scheme uses pixel multiplexing at a single output plane by assigning disjoint subsets of the available output detector pixels to different input planes within the target volume of interest. In the numerical simulations of Fig. 4, we discretize the input volume over which the emitters are distributed by 4 different planes ($P_1$, ..., $P_4$) axially separated by $2.67\lambda$ and the output detector pixels assigned to image these planes are arranged in a rectangular periodic pattern (see Fig. 4a). For this imaging task, the diffractive processor is trained to synthesize a set of 3D PSFs mapping the input points within the target volume onto the corresponding output detector pixels (determined from the arrangement of the output pixels). Figure 4b shows the optimized phase patterns for the diffractive processor designed with $K = 4$ successive surfaces, and the resulting imaging performance with test objects is shown in Fig. 4c. The emission intensities over the input volume are mapped onto the corresponding intensities over a single output/detector plane. Demultiplexing of these raw output pixels at the detector array, assigned to different axial input planes, reveals the emission intensities over each input plane with negligible error. This snapshot 3D imaging of incoherent emitters within an input volume of interest does not require any axial scanning or digital image reconstruction steps[31,32], and it only involves the rearrangement of the output detector pixel values, which achieves axial demultiplexing from a single output image.

The presented 3D information processing framework can also synthesize *spectrally* and *spatially* varying incoherent PSFs, which can enable *snapshot 3D multispectral imaging*. In the numerical simulations illustrated in Fig. 5, three distinct types of emitters, emitting at three different wavelengths (e.g., three different fluorophores) are assumed to be distributed over the input volume; see Fig. 5a. At the output plane corresponding to a detector array, three pixels are assigned to each input voxel: one for each emission wavelength. Demultiplexing of the output pixels assigned to different input planes and emission wavelengths (i.e., *axial and spectral* demultiplexing) reveals the multispectral emission intensities over each input plane; see Fig. 5c. Note that in such a snapshot 3D multispectral imager, the same diffractive network processes all the spectral components, simultaneously outputting the spectral images at the corresponding pixels. The separate demonstrations of 3D imaging at different wavelengths in Fig. 5c further emphasize the negligible cross talk at the output pixels, i.e., when only one type of emitter (corresponding to one emission wavelength) is present, there is negligible leakage signal at the pixels dedicated to the other wavelengths – as desired. These results demonstrate the accurate approximation of a desired set of spatially and spectrally varying 3D incoherent PSFs required for snapshot 3D multispectral imaging of a volume using a single output detector array (without any axial scanning, spectral filters or digital image reconstruction algorithms).

## Discussion

In the numerical analyses presented so far, we assumed that there is no interaction among incoherent emitters of interest. Stated differently, the independent emitters within the target volume at the input do not influence each other, e.g., they do not excite, shadow or block other emitters. If this assumption is violated due to, for example, some emitters scattering or absorbing the emissions of other emitters, the optical system becomes nonlinear from the perspective of input information encoding and cannot be represented by a linear optical forward model ($o = Ai$). In that case, the output 3D intensity patterns, $o$, become a nonlinear function of $i$, and the nature of this nonlinear transformation function depends on the specific distribution and the complex-valued scattering potential of the input object volume, which altogether make the problem significantly more difficult to model due to *emitter-to-emitter coupling*, which is at the heart of nonlinear information encoding. The functional form of such a system becomes object-dependent[33], which means that every unique 3D object topology ($k$) will have a different nonlinear transformation function, i.e., $f_k(i)$, at the output if the emitter-to-emitter coupling is not negligible. For an input volume where the emitters exhibit cross-coupling with each other, the specific 3D topology ($k$) of this volume (dictated by, e.g., the cross-coupling strengths, refractive index distributions, etc.) would change the functional form of the nonlinear transformation, $f_k(i)$; in general, we have $f_k(i) \neq f_m(i)$ for $k \neq m$. These arguments only apply if there is considerable cross-talk among the incoherent emitters of interest; however, weakly scattering and low-density independent emitters located within a uniform and transparent medium would follow the assumptions of our forward model and can be modeled through spatially varying 3D incoherent PSFs represented by our linear system, $o = Ai$.

The large 3D PSF approximation error for $K = 2$ in Fig. 2b emphasizes the importance of structural depth, in terms of the number of successive surfaces, on the performance of diffractive optical processors. Previous works on diffractive processors reported a similar phenomenon, where shallower diffractive networks suffer from relatively larger approximation errors for different desired tasks because of the dominance of the ballistic photons[28,30] at the output plane/volume. With sufficient structural depth, this error is well mitigated, as seen by the small gap between the $K = 4$ and $K = 8$ curves in Fig. 2b.

Regarding the demonstration of snapshot multispectral 3D imaging reported in Fig. 5, it should be noted that all the pixels at the output plane are assumed to be identical with no wavelength specificity, and there is no spectral filtering involved. For a given wavelength, the diffractive processor is optimized to route the photons at the designated pixels and away from the other pixels (eliminating spectral cross-talk), as shown in the raw 'Output intensity' and 'Axially and spectrally demultiplexed output images' in Fig. 5c. Accordingly, when independent emitters of different types (i.e., emitting at different wavelengths) are present in the same input volume, such a snapshot 3D multispectral imager simultaneously processes all the spectral components of the 3D emitters within the input volume to output the spectral images at the corresponding pixels of the output detector array.

Another important feature of the presented diffractive processors is that their design, i.e., the diffractive layers and the corresponding features optimized for a desired 3D optical information processing task, can be translated to different parts of the electromagnetic spectrum without redesigning the layers due to the scale invariance of the underlying formalism based on Maxwell's equations. Furthermore, the spectral engineering of the spatially varying 3D PSFs that is achieved using diffractive processors (e.g., for the snapshot multispectral 3D imaging reported in Fig. 5) ) does *not* rely on the specific properties or optimization of the material dispersion; the dispersion resulting from free-space propagation of optical waves suffices in the spectral engineering of spatially varying 3D PSFs even if the material dispersion is negligible with a relatively flat refractive index as a function of wavelength[29]. Therefore, the conclusions of our analyses are broadly applicable to various materials of interest that operate at different parts of the electromagnetic spectrum without the need for redesigning the task-specific optimized diffractive features.

We believe that the results presented here lay the groundwork for 3D optical information processing systems with incoherent diffractive processors. As demonstrated through our numerical analyses, diffractive processors can be optimized to process 3D spatial information at multiple wavelengths through desired sets of spatially and spectrally programmed incoherent PSFs, covering various applications in imaging and sensing with unique capabilities beyond traditional spatially invariant free-space optics-based processors.

## Supplementary Information:

Methods Section

- Optical forward model of spatially incoherent diffractive networks
- Intensity linear transformations by a spatially incoherent diffractive optical processor
- Optimization of spatially incoherent diffractive optical processors

# Figures

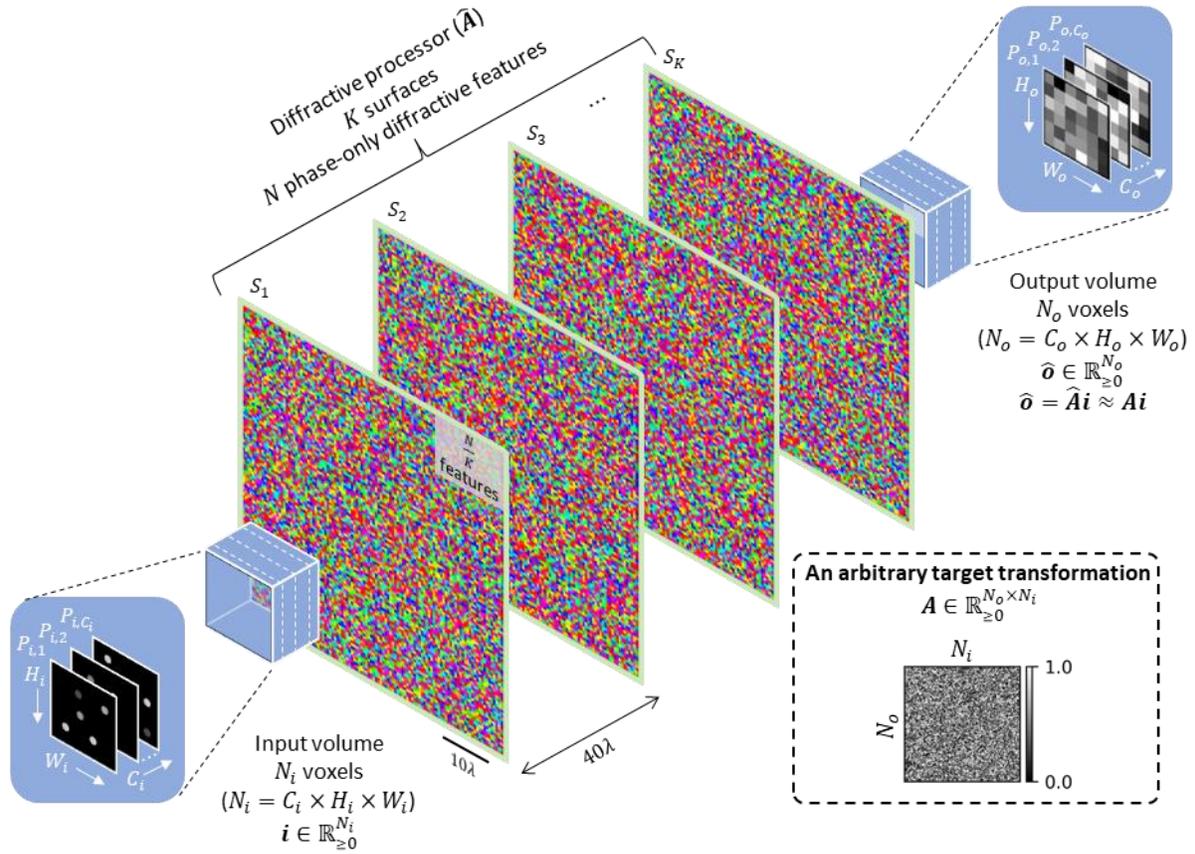

**Fig. 1** 3D PSF engineering using a spatially incoherent diffractive optical processor. The diffractive processor, equipped with $N$ optimizable phase-only features that are distributed over $K$ surfaces, performs a linear transformation $\widehat{A} \approx A$ on the 3D input intensity $i$ to create the 3D output intensity distribution $\widehat{o} = \widehat{A}i$; here $A \in \mathbb{R}_{\geq 0}^{N_o \times N_i}$ is an arbitrarily defined target transformation, the columns of which represent the target (desired) 3D PSFs that are spatially varying. $N_i$ and $N_o$ are the numbers of diffraction-limited voxels within the input and output volumes, respectively.

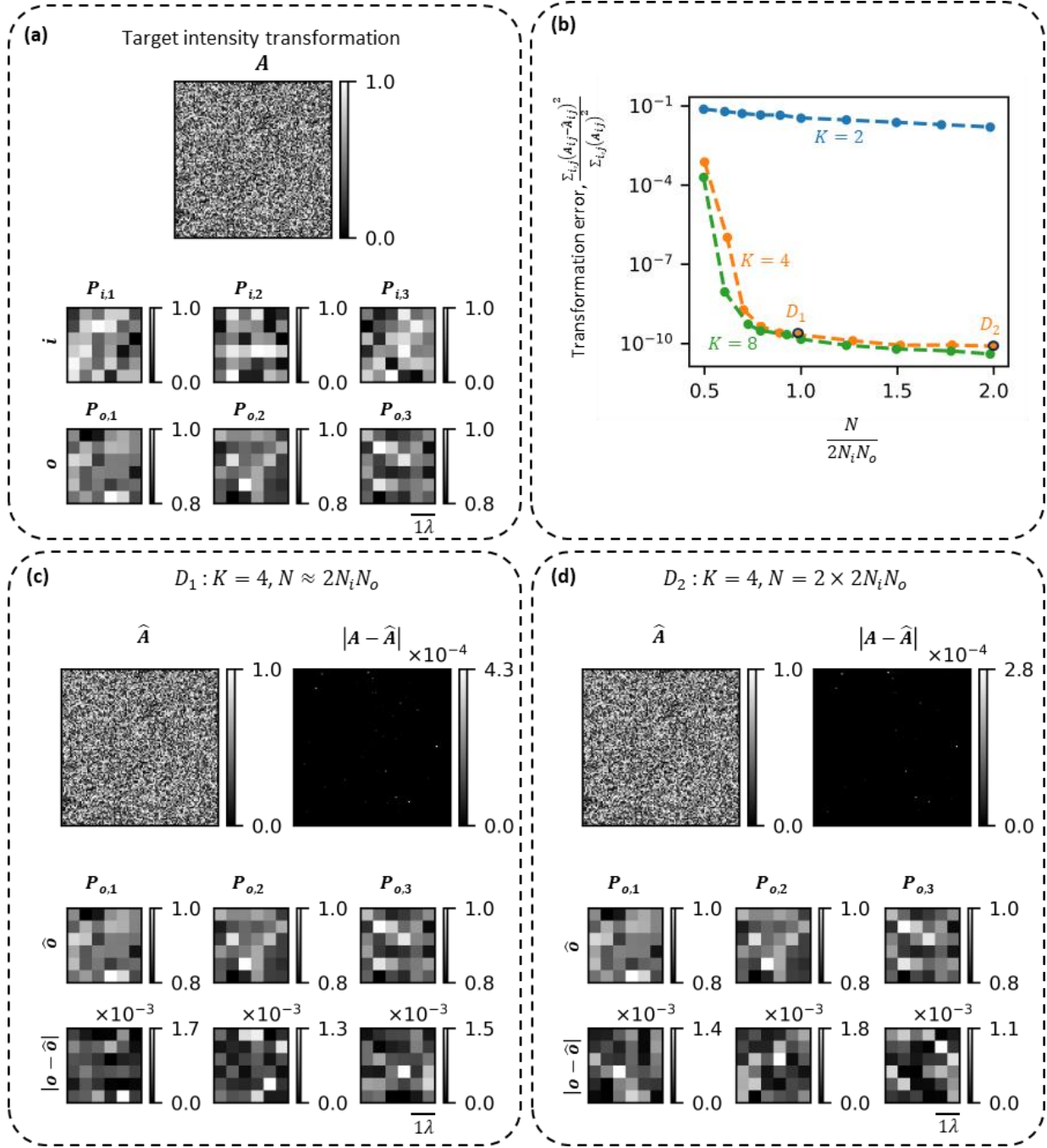

**Fig. 2** 3D PSF-approximation error as a function of the number ($N$) of optimizable phase-only diffractive features. (a) Top: Target intensity linear transformation $A$, representing the target spatially varying 3D PSFs. Bottom: An example of 3D input intensity $i$ and the corresponding (target) 3D output intensity $o$. (b) Error in the all-optical approximation of the target linear transformation, as a function of $N$ where the $N$ optimizable phase-only features are distributed over $K$ surfaces. (c) 3D PSF-approximation performance of a diffractive processor with $K = 4$, $N \approx 2N_i N_o$, labeled as $D_1$ in Fig. 2b. Top: The all-optical transformation $\widehat{A}$, together with the elementwise absolute error, which is negligible. Bottom: The 3D output intensity $\widehat{o}$ corresponding to the input intensity $i$ depicted in Fig. 2a, along with the elementwise absolute error, which is negligible. (d) 3D PSF-approximation performance of a diffractive processor with $K = 4$, $N = 2 \times 2N_iN_o$, labeled as $D_2$ in Fig. 2b. Top: The all-optical transformation $\widehat{A}$, together with the elementwise absolute error, which is negligible. Bottom: The 3D output intensity $\widehat{o}$ corresponding to the input intensity $i$ depicted in Fig. 2a, along with the elementwise absolute error, which is negligible.

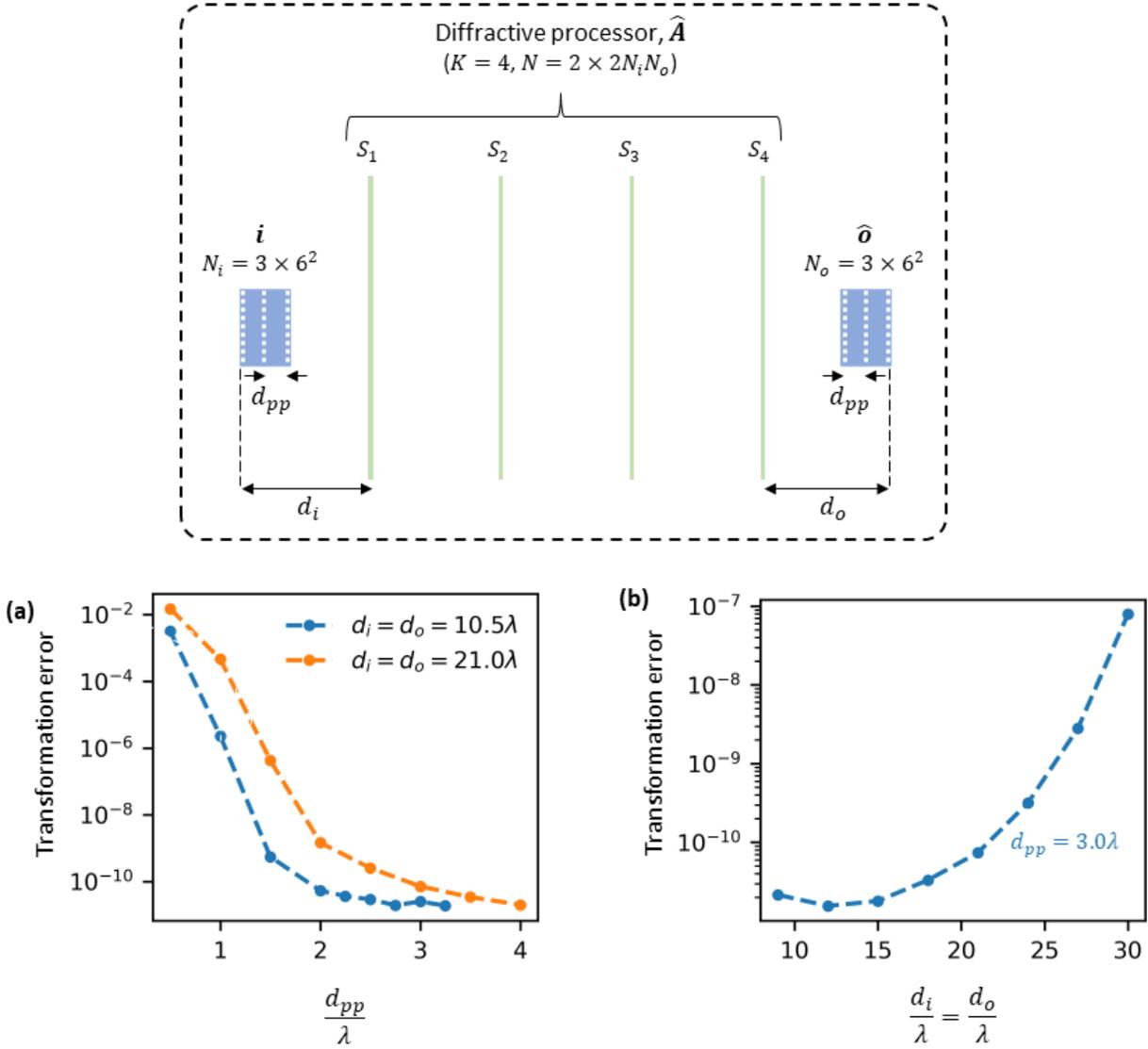

**Fig. 3** Effect of diffraction limit on the 3D PSF-approximation error. (a) 3D PSF-approximation error as a function of plane-to-plane distance $d_{pp}$ within the input and output volumes, while the distances of the first input plane ($d_i$) and the last output plane ($d_o$) from the diffractive processor are kept constant. The two curves correspond to two different values of $d_i = d_o$. (b) 3D PSF-approximation error as a function of $d_i = d_o$, while the plane-to-plane distance $d_{pp}$ is kept constant.

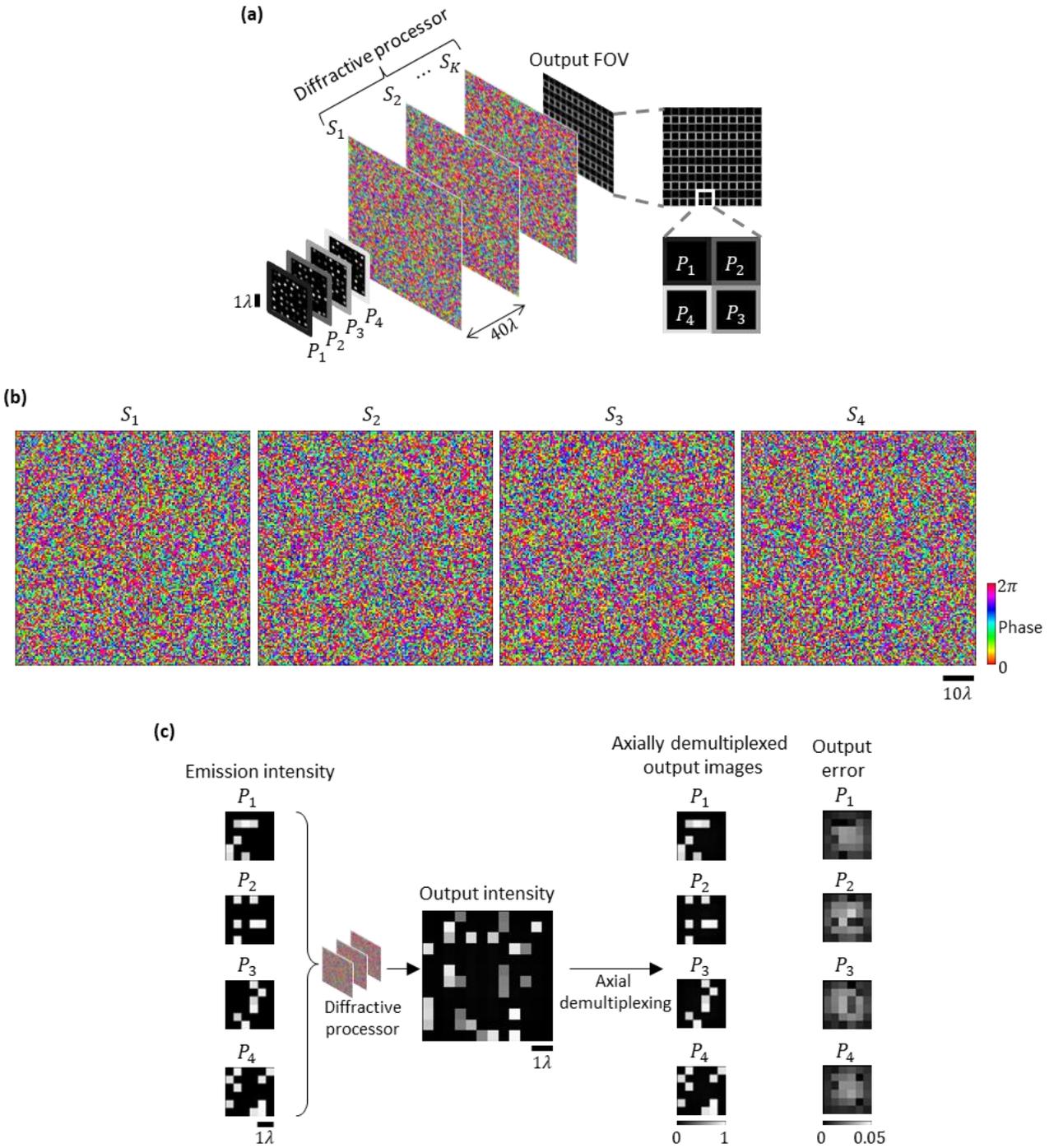

**Fig. 4** Snapshot 3D imaging using a spatially incoherent diffractive optical processor. (a) The pixels on a single output plane are grouped into 'superpixels'. Each constituent pixel of an output superpixel corresponds to one input plane, and the number of superpixels on the output plane is equal to the number of pixels on each input plane. The shaded outlines around the pixels denote the assignment to the respective input planes. The intensities at the set of output pixels assigned to an input plane constitute the image of that input plane. (b) The optimized phase profiles of the $K = 4$ surfaces of a diffractive processor, designed for snapshot 3D imaging over four input planes. (c) Each input plane is discretized with $6 \times 6$ diffraction-limited pixels, with a distinct emitter configuration in each plane. An example input intensity pattern at the input volume and the corresponding output image, together with the elementwise absolute error, which is negligible.

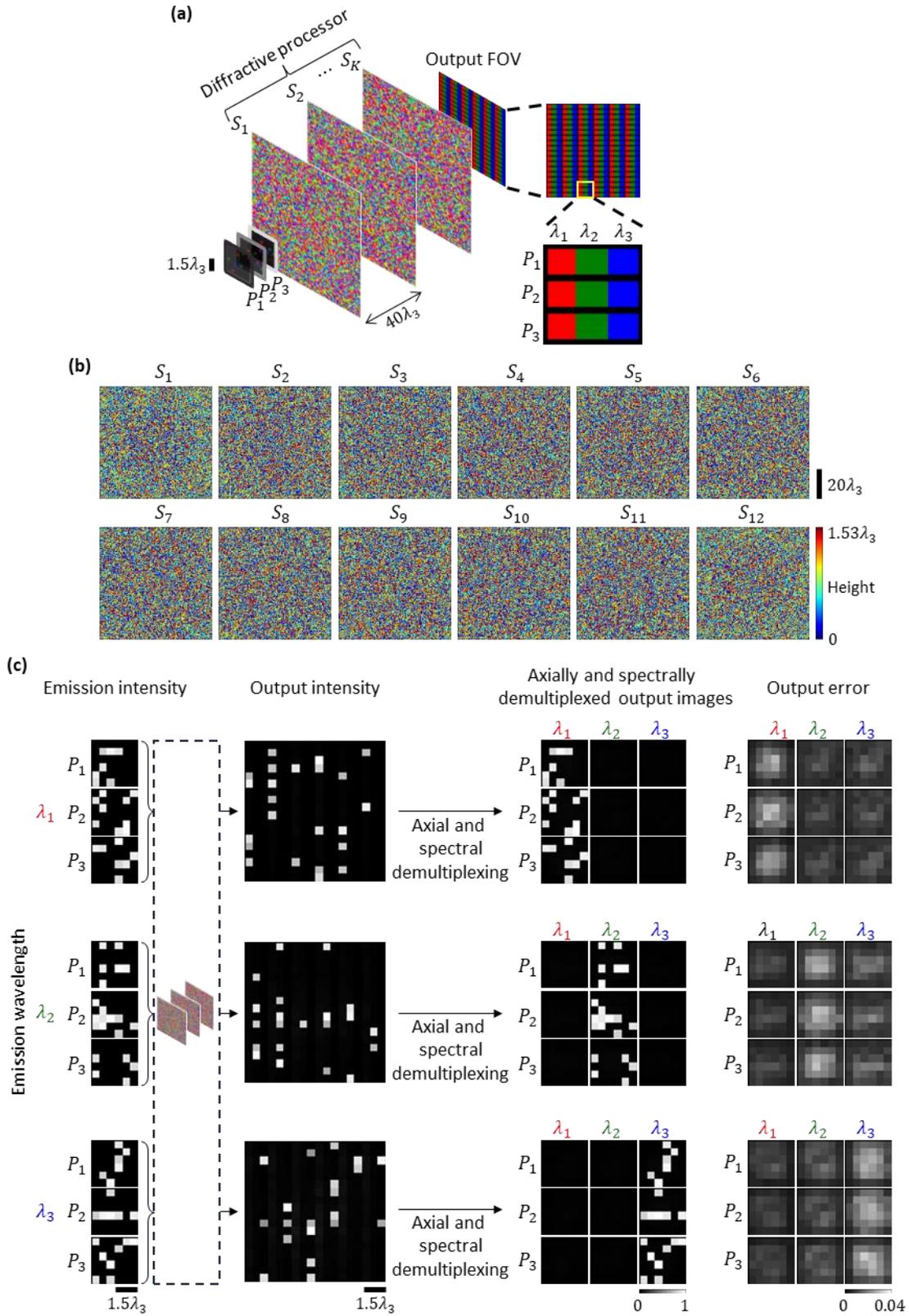

**Fig. 5** Snapshot multispectral 3D imaging using a spatially incoherent diffractive optical processor. (a)

The pixels on a single output plane are grouped into 'superpixels'. Each constituent pixel of a superpixel corresponds to one input plane and one wavelength, and the number of superpixels on the output plane is equal to the number of pixels on each input plane. The fill colors denote the assignment to the respective emission wavelengths. The intensities at the set of output pixels assigned to an input plane and a wavelength constitute the spectral image of that corresponding input plane. (b) The optimized thickness profiles of the $K = 12$ surfaces of a diffractive processor, designed for snapshot 3D imaging over 3 input planes at 3 wavelengths. Each plane is discretized with 6×6 pixels. (c) Example input intensity patterns at different wavelengths and the corresponding output images, together with the elementwise absolute error, which is negligible.